\documentclass[review,number,sort&compress,5p]{elsarticle}
\usepackage{lineno}
\usepackage{color}
\usepackage{bm}        
\usepackage{amssymb}   
\usepackage{amsmath}
\usepackage{longtable}
\usepackage{dcolumn}
\usepackage{multirow}
\usepackage{verbatim}

\journal{Nuclear Instruments and Methods in Physics Research Section A}

\begin{document}
\begin{frontmatter}

\title{The TITAN in-trap decay spectroscopy facility at TRIUMF}

\author[TRI,SFU]{K.G.~Leach\corref{cor1}}\ead{kleach@triumf.ca}
\author[TRI]{A.~Grossheim}
\author[TRI,MUN]{A.~Lennarz}
\author[TRI,TUM,STA]{T.~Brunner}
\author[HEI]{J.R.~Crespo~L\'opez-Urrutia}
\author[TRI,UBC]{A.T.~Gallant}
\author[TRI]{M.~Good}
\author[TRI,HEI]{R.~Klawitter}
\author[TRI]{A.A.~Kwiatkowski}
\author[SFU]{T.~Ma}
\author[TRI,UBC]{T.D.~Macdonald}
\author[SFU]{S.~Seeraji}
\author[TRI]{M.C.~Simon\fnref{label1}}
\author[SFU]{C.~Andreoiu}
\author[TRI,UBC]{J.~Dilling}
\author[MUN]{D.~Frekers}

\cortext[cor1]{Corresponding author}
\fntext[label1]{Present address: Stefan Meyer Institute for Subatomic Physics, 1090 Vienna, Austria}

\address[TRI]{TRIUMF, 4004 Wesbrook Mall, Vancouver, British Columbia V6T 2A3, Canada}
\address[SFU]{Department of Chemistry, Simon Fraser University, Burnaby, British Columbia V5A 1S6, Canada}
\address[MUN]{Institut f\"ur Kernphysik, Westf\"alische-Wilhelms-Universit\"at M\"unster, D-48149 M\"unster, Germany}
\address[TUM]{Physik Department E12, Technische Universit\"at M\"unchen, D-85748 Garching, Germany}
\address[STA]{Department of Physics, Stanford University, Stanford, CA 94305, USA}
\address[HEI]{Max-Planck-Institut f\"ur Kernphysik, D-69117 Heidelberg, Germany}
\address[UBC]{Department of Physics and Astronomy, University of British Columbia, Vancouver, BC, V6T 1Z1, Canada}

\begin{abstract}
This article presents an upgraded in-trap decay spectroscopy apparatus which has been developed and constructed for use with TRIUMF's Ion Trap for Atomic and Nuclear science (TITAN).  This device consists of an open-access electron-beam ion-trap (EBIT), which is surrounded radially by seven low-energy planar Si(Li) detectors.  The environment of the EBIT allows for the detection of low-energy photons by providing backing-free storage of the radioactive ions, while guiding charged decay particles away from the trap centre via the strong (up to 6~T) magnetic field.  In addition to excellent ion confinement and storage, the EBIT also provides a venue for performing decay spectroscopy on highly-charged radioactive ions.  Recent technical advancements have been able to provide a significant increase in sensitivity for low-energy photon detection, towards the goal of measuring weak electron-capture branching ratios of the intermediate nuclei in the two-neutrino double beta ($2\nu\beta\beta$) decay process.  The design, development, and commissioning of this apparatus are presented together with the main physics objectives.  The future of the device and experimental technique are discussed.
\end{abstract}


\end{frontmatter}


\section{Introduction}
\subsection{High sensitivity decay spectroscopy}
The characterization of radioactive decay via photon detection is a key measurement method and is among the primary experimental techniques currently employed in nuclear physics~\cite{KNOLL}.  With the advancement of rare-isotope beam (RIB) facilities worldwide~\cite{Blu13}, access to increasingly exotic radioactive nuclei has become possible, allowing for a variety of decay experiments on short- and long-lived nuclei.  Modern decay spectroscopy devices employ multiple detection systems for both charged particles and photons to further increase the sensitivity of the experiment, thus allowing for the observation of weak signals~\cite{Dun13}.  The reduction of photon backgrounds is at the forefront of these efforts, and requires a high level of control over the decay environment which can be provided using ion traps~\cite{Bla13}.

The concept of observing decays from trapped radioactive nuclei has been employed for years, most notably using magneto-optical traps and Paul traps, where charged particles and daughter recoils are detected to provide direct and indirect information about neutrinos~\cite{Gor05,Vet08,Fle08,Koz08,Li13}, electrons~\cite{Cou12}, and neutrons~\cite{Yee13}.  More recently, Penning traps have been considered to provide control over the decay environment~\cite{Ris07,Ett09,Meh12,Bru13,Web13,Rin14,Len14,Len14b,Asn14}, where charged particles are guided along strong magnetic-field lines.  Therefore, further extension of this concept may be possible for performing high-sensitivity decay-spectroscopy measurements.

\subsection{\label{NME}Nuclear matrix elements for $\beta\beta$ decay}
Recent evidence that neutrinos have mass has generated a great deal of interest in exotic nuclear decay modes~\cite{Cir13,Avi08}.  As a part of these studies, searches for the neutrinoless ($0\nu$) mode of double beta ($\beta\beta$) decay is among the most relevant since it violates lepton-number conservation and would establish the neutrino as a Majorana particle~\cite{Sch82,Tak84}.  If this decay mode is observed, the effective Majorana mass of the neutrino, $\langle m_{\beta\beta}\rangle$, can be deduced from $0\nu\beta\beta$ measurements,
\begin{eqnarray}
\label{0nubb}
(T^{0\nu}_{1/2})^{-1}=G^{0\nu}(Q,Z)|M^{0\nu}|^2\langle m_{\beta\beta}\rangle^2,
\end{eqnarray}
\noindent
where $T^{0\nu}_{1/2}$ is the observed half-life of the $0\nu\beta\beta$ decay and $G^{0\nu}(Q,Z)$ is the phase-space factor.  The term $M^{0\nu}$ is the nuclear matrix element (NME) connecting the initial and final $0^+$ states, which results entirely from theoretical calculations.  The calculation of $\beta\beta$-decay NMEs is the source of current theoretical efforts and include several different model descriptions.  The accuracy and precision from Eqn.~\ref{0nubb} is limited by the ability to calculate the NMEs, and any uncertainty in the calculations are directly translated to $\langle m_{\beta\beta}\rangle$.  Therefore, constraints on these calculations are required from detailed experimental data.

Typically, the NME calculations are benchmarked to two-neutrino ($2\nu$) $\beta\beta$ data~\cite{Bar10} (a process allowed by the Standard Model) where the decay path proceeds through $1^+$ states in the odd-odd intermediate nucleus~\cite{Fre07}.  Therefore, measurements of the $\beta^-$ and electron-capture (EC) branching ratios of the intermediate nuclei in the $2\nu\beta\beta$-decay process are directly relevant for capturing the nuclear-physics information required in the calculation of $M^{2\nu}$.  The EC transitions are several orders of magnitude weaker than the dominant $\beta^-$ decays from the same parent nucleus, making them difficult to detect.  Due to the weak nature of these decay branches, these studies require intense RIBs and low-background, high-sensitivity decay spectroscopy tools~\cite{Fre07,Bru13}.

\subsection{TITAN at TRIUMF-ISAC}
\begin{figure}[t!]
\begin{center}\includegraphics[width=\linewidth]{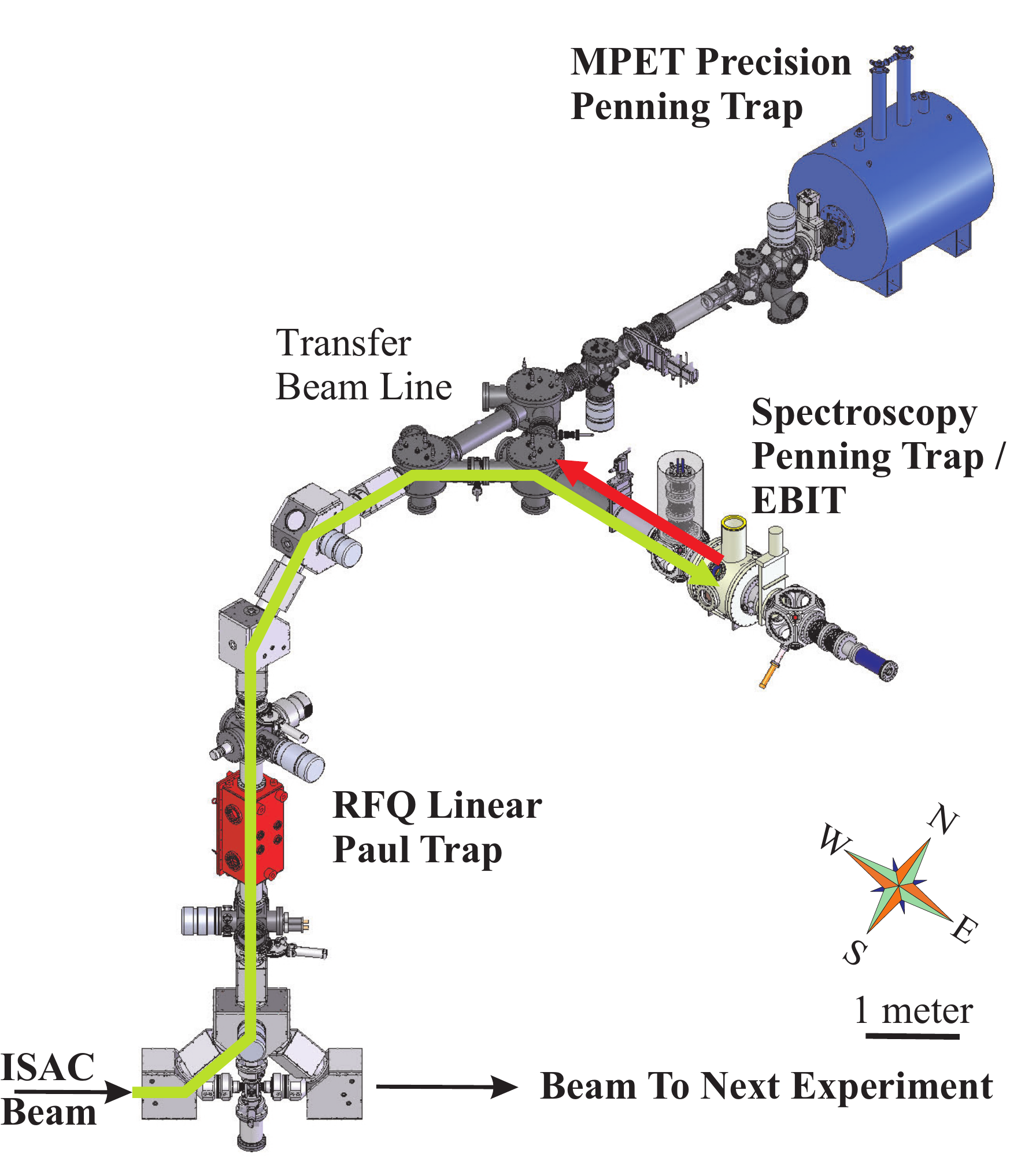}\end{center}
\caption{\label{TITAN}A schematic view of the TITAN facility at TRIUMF.  For decay spectroscopy experiments, the cooled/bunched ions are extracted from the RFQ and injected as singly charged ions to the EBIT (path shown in green), where they are stored and charge-bred.  The ion-bunch is subsequently extracted, and dumped downstream away from the photon detectors (path shown in red).}
\end{figure}
The Isotope Separator and Accelerator (ISAC) facility~\cite{Bri97,Dom00} at TRIUMF in Vancouver, Canada, employs a high-intensity (up to 100 $\mu$A) beam of 500 MeV protons to produce RIBs using the isotope separation on-line (ISOL) technique~\cite{Blu13,Kos02}.  ISAC is currently able to provide a wide variety of RIBs through the use of several different production target and ion-source combinations, including the recent use of uranium-carbide ($UC_x$) targets~\cite{Dil14}.  Following the in-target production and ionization, the ions are mass separated before being delivered to the experimental hall. The mass-selected, continuous beam of radioactive, singly charged ions (SCIs) is delivered at low energies ($<60$~keV) to a suite of experimental facilities for both cooled- and stopped-beam experiments~\cite{Dil14}, where TRIUMF's Ion Trap for Atomic and Nuclear Science (TITAN)~\cite{Dil03,Dil06} is located.  The TITAN system consists of three ion traps:
\begin{enumerate} 
\item A radio-frequency quadrupole (RFQ) linear Paul trap~\cite{Smi06,Bru12} for buffer-gas cooling and bunching the low-energy ion beam,
\item A 3.7~T, high-precision mass-measurement Penning trap (MPET)~\cite{Bro12}, and
\item An electron-beam ion trap (EBIT) which is used to create highly charged ions (HCIs)~\cite{Lap10}, and for performing decay spectroscopy on trapped radioactive nuclei.
\end{enumerate}
A schematic view of the TITAN facility at TRIUMF-ISAC is shown in Fig.~\ref{TITAN}, along with the ion path for typical decay-spectroscopy experiments.

\subsection{In-Trap Decay Spectroscopy with TITAN}
\begin{figure}[t]
\includegraphics[width=\linewidth]{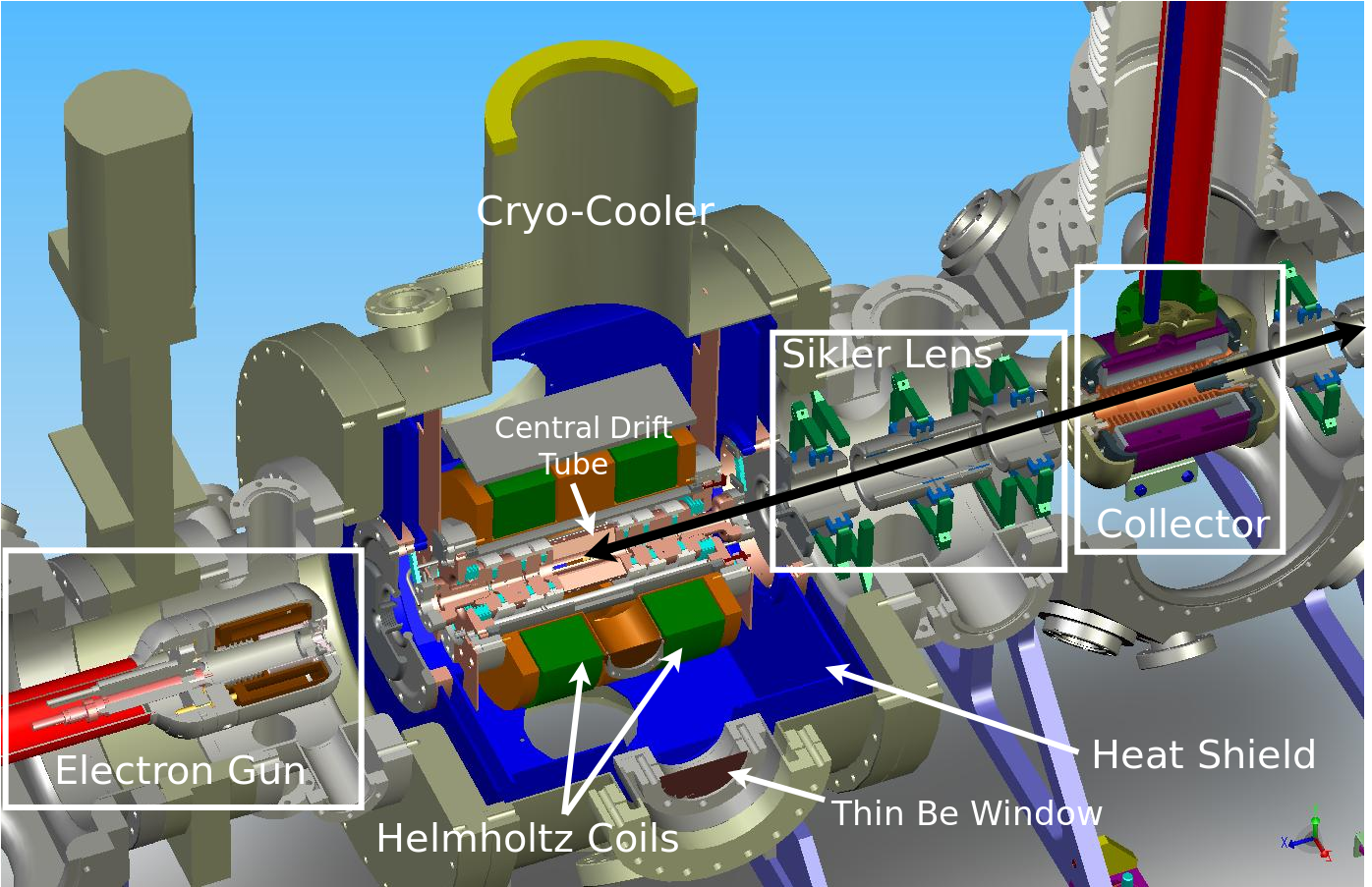}
\caption{\label{EBIT1}A technical depiction of the TITAN EBIT.  The ion-bunch trajectories during operation are schematically depicted here by the black double-arrow.  The individual components are discussed further in the text.}
\end{figure}
The prospect of performing decay spectroscopy with TITAN was first presented in Refs.~\cite{Ett09,Bru13}. In these measurements, a low-energy germanium (LEGe) detector was placed in the EBIT for photon counting, and no electron beam was employed for ion confinement of charge breeding.  In this mode of operation, the EBIT effectively serves as a cylindrical Penning trap.  The results from these measurements demonstrated that ions could be injected, stored, and extracted from the EBIT for the purpose of decay spectroscopy, however storage times were limited to tens of ms due to losses at the trap center. Additionally, the in-trap losses meant that information regarding the precise location of where the decays were occurring was lost, and thus a determination of the photon detection efficiency was not possible.  Since the primary science goal of this apparatus is the characterization of weak decay branches ($10^{-3}$-$10^{-5}$), an improvement of the experiment was required, and new techniques were developed.

This article presents a significant upgrade to the apparatus, and addresses the above deficiencies towards the goal of high-sensitivity in-trap decay spectroscopy.  These improvements include: a new trapping mechanism, different photon detectors, improved ion-bunch manipulation, a superior data-acquisition system, and improved environmental monitoring and control.

\section{The Decay Spectroscopy Trap}
\begin{figure*}[t!]
\centering
\includegraphics[width=\linewidth]{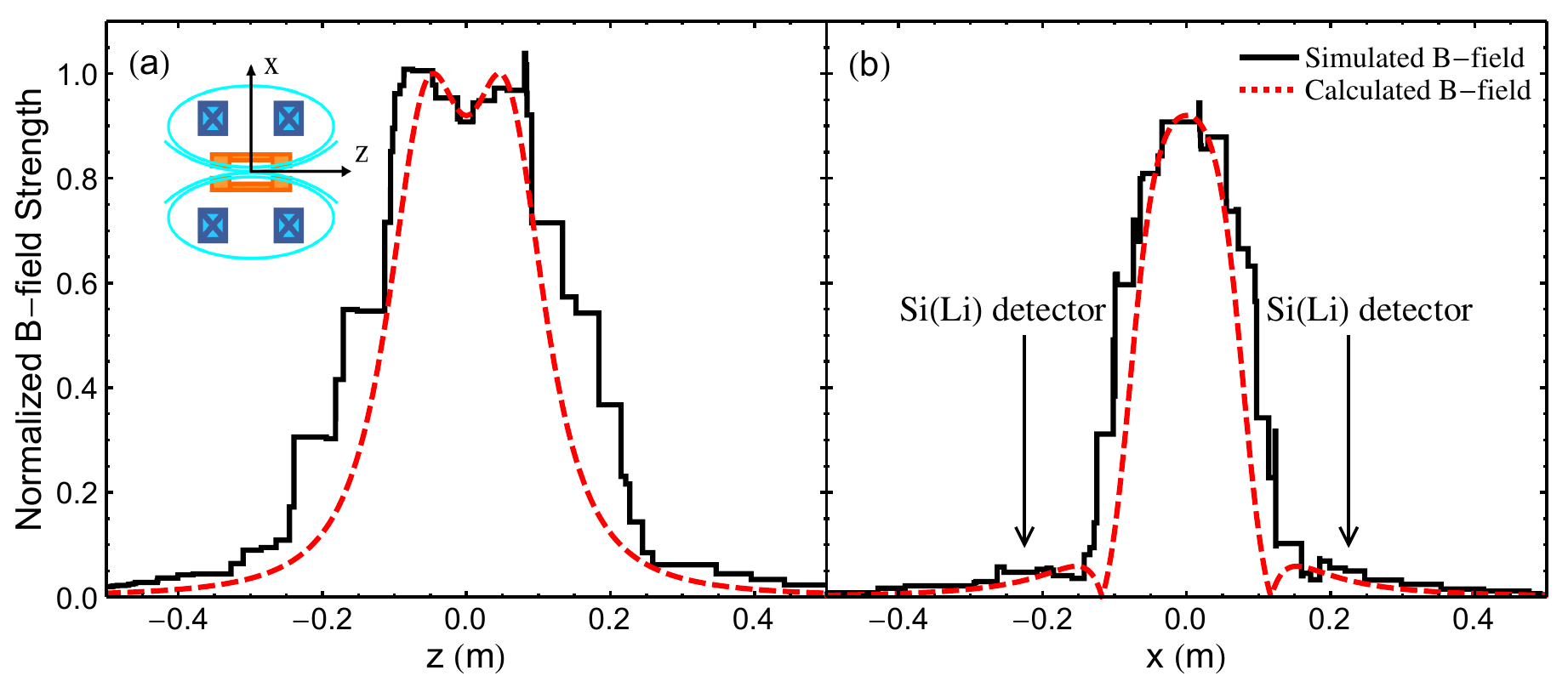}
\caption{\label{BfieldFig}Simulated~\cite{COMSOL} and calculated~\cite{Sto00} magnetic field strengths (a) along and (b) perpendicular to the trap axis, for the Helmholtz-like configuration used in the EBIT~\cite{Lap10}.  The relative magnetic field strengths are normalized to the maximum value, which occurs along the trap axis at the coils.  The arrows in panel (b) indicate the radial location of the photon detectors used for decay spectroscopy.}
\end{figure*}
The TITAN EBIT (Fig.~\ref{EBIT1}) is composed of an up-to $500$~mA electron gun\footnote{An upgrade for the electron gun is planned, which will allow for beam currents of up to 5~A.}, a cold drift-tube assembly which is thermally coupled to a superconducting magnet, and an electron collector.  The drift-tube assembly is conically shaped which improves the trapping profile while retaining a large geometric acceptance for the incoming ions.  The 8-fold radially segmented central electrode forms the potential where the ions are stored during the trapping cycle, and has an inner radius of 7.0~mm~\cite{Lap10}.  The trapped ions are axially confined by an electrostatic square-well potential formed by applying voltage to the drift tubes.  Radial confinement is provided by both the electron-beam space-charge potential and magnetic field.  The up-to 6~Tesla magnetic field is produced by two superconducting Nb$_3$Sn coils in a Helmholtz-like configuration.  At the trap center, the field is reduced by 8\% from this configuration, as shown in Fig.~\ref{BfieldFig}, creating a magnetic bottle.  The radial confinement provides a spatial profile for the ions which is approximately equal to the radial extent of the electron beam.

\subsection{Ion Storage}
The total ion trapping capacity of an electron-beam device is roughly determined by the fact that a significant fraction of the electron beam negative space charge can be compensated by the trapped positive ions~\cite{Pen91}.  The number of negative elementary charges within the central trap region depends linearly on the electron beam current, and inversely on the square root of its energy.  For the TITAN EBIT, with an up-to 500 mA, 10 keV electron beam, the trapping region contains roughly $10^9$ electrons.  For an average ion charge state of $q\approx30^+$, this implies that roughly $10^7$ ions can be confined.  This relatively large value is only possible due to the negative space charge of the beam, which counterbalances the mutual ion repulsion acting in a pure Penning configuration.  Furthermore, under these operating parameters, the radial trapping potential for a positively charged ion of $q=30^+$ is on the order of 10 keV.  Under these conditions, ion losses are expected to be very small, and the cycling of the charge state of the ion due to successive ionization and recombination processes does not affect the ion inventory. In principle, the axial evaporation, which is controlled by voltages applied to the drift tubes surrounding the central region, is the essential loss mechanism.  Since Ba and W naturally accumulate in the trap due to their emission from the cathode material, elements lighter than $Z=50$ may suffer stronger evaporative losses.  An evaluation of possible in-trap losses with the TITAN EBIT are discussed further in Section~\ref{storagelosses}.

\subsection{\label{traj}Trajectories of Charged Decay Products}
\begin{figure}[t]
\includegraphics[width=\linewidth]{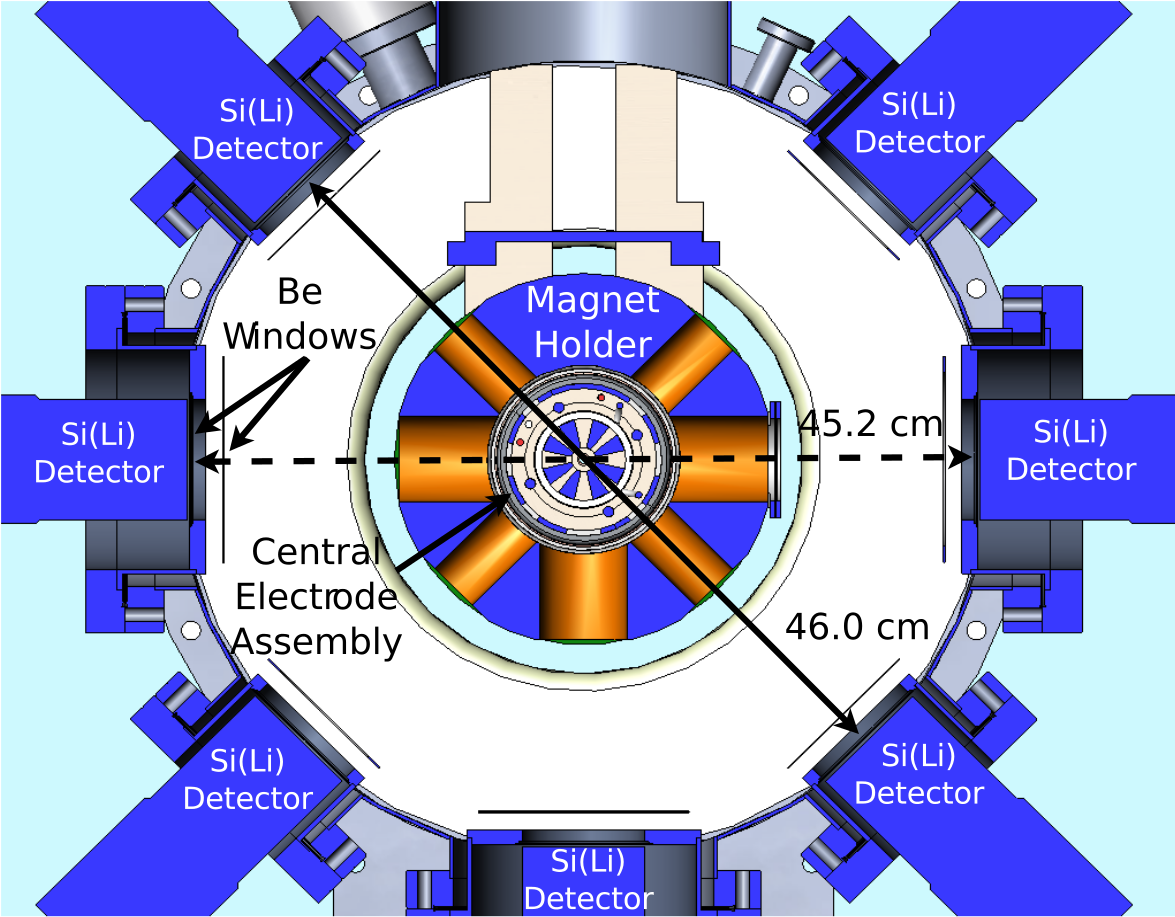}
\caption{\label{trap_angles}A design drawing cross-sectional view of the EBIT.  The magnet housing and central-electrode assembly shown in Fig.~\ref{trap_interior} are at the center of the trap, and are surrounded by the vacuum vessel.  Seven access ports are cut into the outer housing of the EBIT, at two different distances from the center, and each holds one Si(Li) detector.  The dimensions labelled represent the distances between the front faces of Si(Li) detectors on opposite sides of the trap.}
\end{figure}
The primary advantage of performing decay spectroscopy in a high magnetic field environment stems from the removal of light, charged decay products which generate large background (ie. $\beta^{+/-}$, p, $\alpha$, etc.).  For the physics cases outlined in Section~\ref{NME} and Ref.~\cite{Fre07}, a reduction of large decay-electron backgrounds from $\beta^-$ decay is required.  In the EBIT, $\beta$-particles that are emitted from the confined ion bunch follow the magnetic field lines, and are guided away from the trap volume along its axis.  From SIMION~\cite{Dah00} simulations at $\vec{B}$-field strengths of 4, 5, and 6\,T, the fraction of decay electrons that escape is roughly 77\%.  This process requires that the emission angle of the $\beta$-particle must be less than the critical trapping angle.  This critical angle depends on the ratio between the magnetic field at the electron origin and the maximal magnetic field.  However, nearly 100\% of the charged decay products are eliminated from the trap, since the $\beta$-particles have a high probability to both Coulomb scatter with the ion cloud, and cool via synchrotron radiation.  The increase in scattering probability results from the high cyclotron frequency for electrons in a high-field environment~\cite{Asn14}.  Both of the above processes cause the decay products to drop below the critical angle and immediately escape the trap volume along the beam axis.  These studies are particularly important within the context of $\beta^+$ decay, as the complete removal of decay positrons serves to suppress 511~keV annihilation radiation~\cite{Len14,Len14b}.
\begin{figure}[t]
\includegraphics[width=\linewidth]{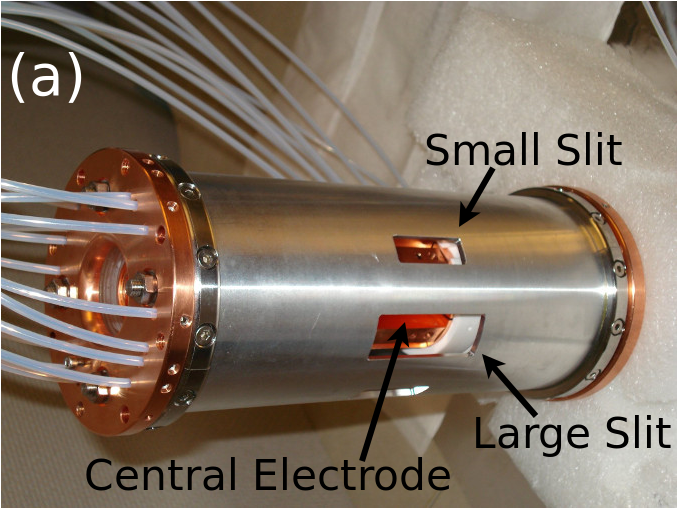}\\
\includegraphics[width=\linewidth]{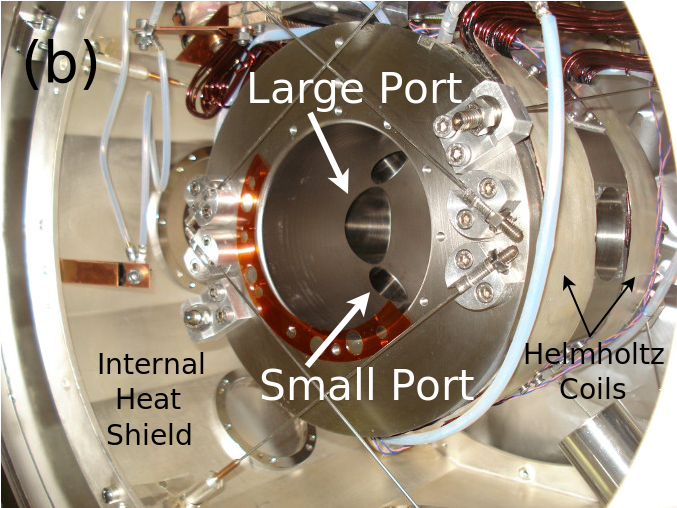}
\caption{\label{trap_interior}Photographs of the components that comprise the TITAN EBIT interior.  Displayed are: (a) the central electrode (copper) and the housing cylinder (aluminum) that sit inside (b) the magnet coil holder.  The solid-angle acceptance for photon detection from in-trap decays are limited by the slits in the electrode housing, shown in panel (a).}
\end{figure}

\subsection{Trap Access}
The EBIT features seven external ports of two different sizes, separated by $45^\circ$ from each other (Fig.~\ref{trap_angles}), each with a 35.0~mm radius opening.  These ports are covered with 0.25~mm thick, $>99\%$ pure, pinhole-free Be windows to provide vacuum isolation for the ultra-high vacuum (UHV) environment of the EBIT, which is better than $10^{-11}$ Torr.  A separate Be foil (0.08~mm thick) is located on the internal heat shield of the trap.  The access ports corresponding to the large slits in the electrode-housing cylinder (Fig.~\ref{trap_interior}(a)) are located at $90^\circ$, $180^\circ$, and $270^\circ$ relative to the cryo-cooler at the top of the magnet housing.  Due to the design of the EBIT, the detectors located in the small ports are slightly further from the trap centre (230~mm), as opposed to those located in the large ports (226~mm) (see Fig.~\ref{trap_angles}).
\begin{figure*}[t]
\includegraphics[width=0.52\linewidth]{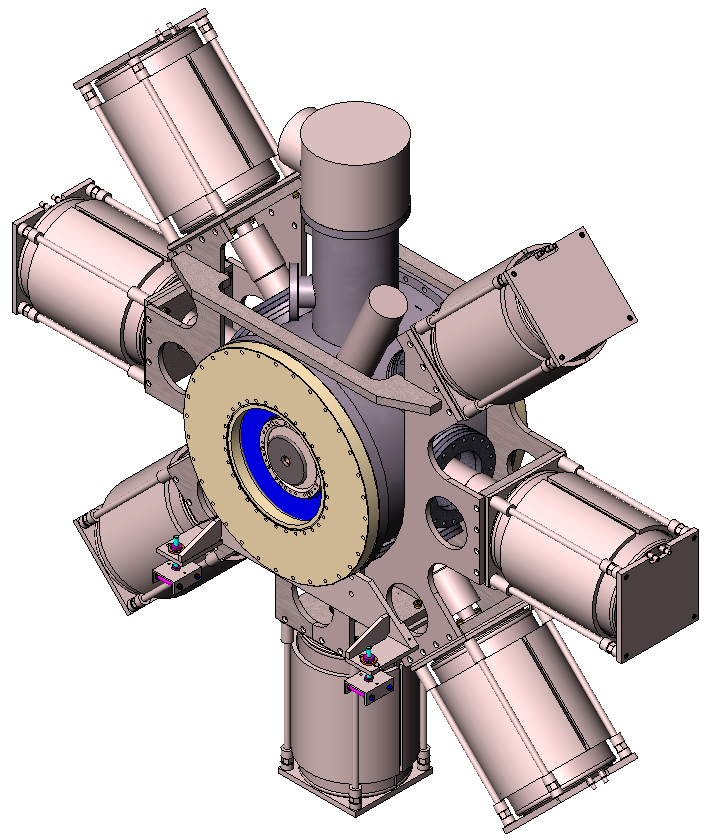}\includegraphics[width=0.48\linewidth]{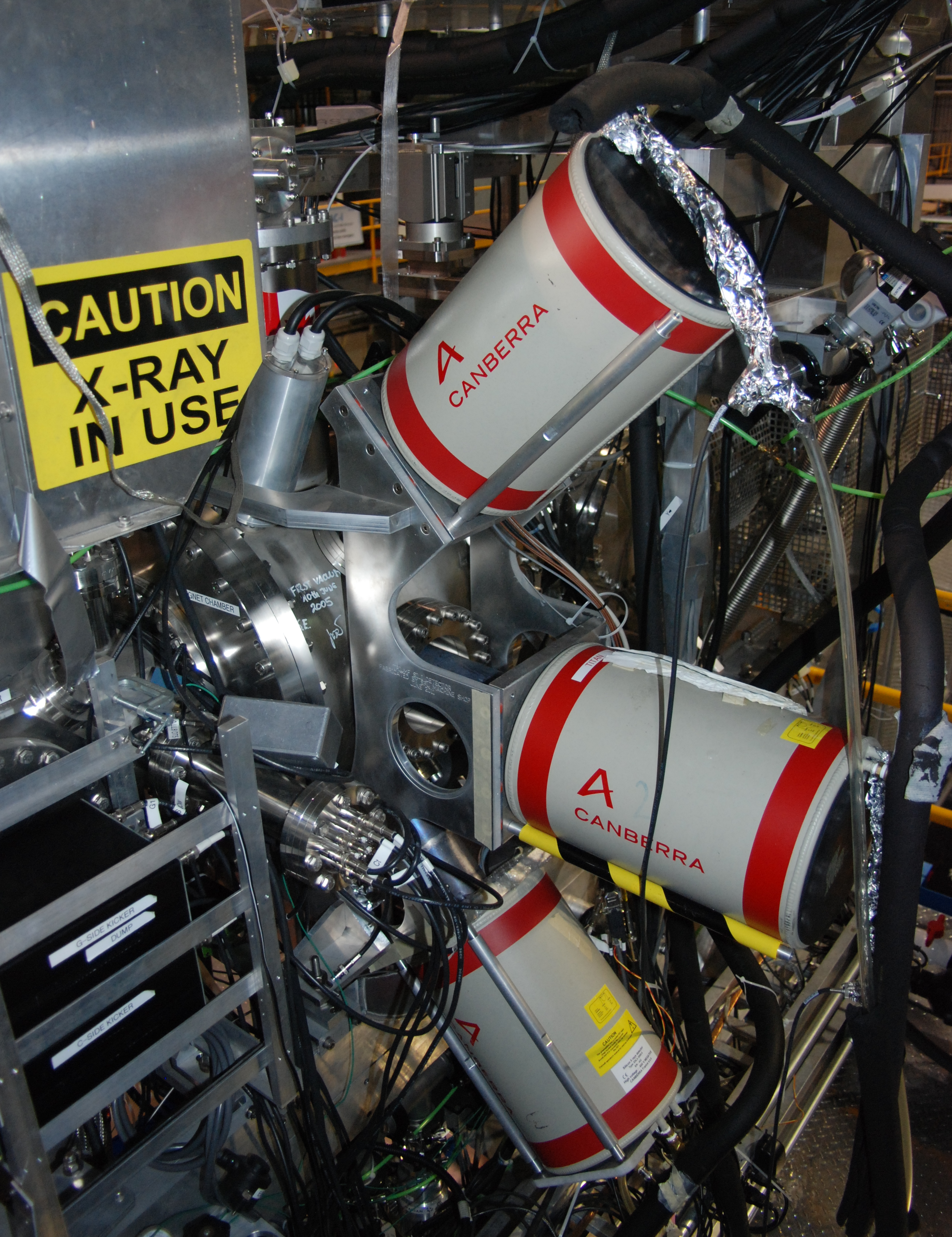}
\caption{\label{TITANEC}(left) A technical drawing of the TITAN decay spectroscopy setup. (right) A photograph of three of the seven Si(Li) detectors that surround the southern hemisphere of the EBIT.  The $e$-collector is at the left of the image, where ion bunches from the RFQ are injected and extracted from the trap.}
\end{figure*}

\subsection{EBIT Operation}
The EBIT is operated in a cycling mode which typically consists of three parts: injection, storage/trapping, and extraction.  The cycles are controlled by logic signals that are sent to the drift-tube electrodes by a programmable pulse-generator (PPG).  For decay spectroscopy experiments, the cycles are optimized to increase the signal-to-background ratio for the species of interest, and thus trapping portions of the cycle can last anywhere from a few seconds to minutes.  These values are typically $10^3$ to $10^4$ times longer than is usually employed for charge-breeding related to the mass-measurement program with TITAN.

\section{\label{photon}Planar Si(Li) Detectors}
Each of the seven access ports around the EBIT house a lithium-drifted silicon detector (Si(Li)) (Fig.~\ref{TITANEC}), which has good resolution and high efficiency at low photon energies ($<50$~keV)~\cite{KNOLL}.  These detectors were chosen over high-purity germanium (HPGe) crystals due to their decreased X-ray escape peak intensity\footnote{Roughly four orders of magnitude at 20~keV~\cite{KNOLL}.} and the prospect of performing a high-sensitivity X-ray measurement on $^{76}$Ge~\cite{Fre07}.

\subsection{Design specifications}
The detectors were designed and constructed by Canberra-France, and each contains a 5~mm thick Si(Li) crystal with a $<0.2~\mu$m dead-layer and 2000 mm$^2$ active surface area.  Each crystal is located 7~mm from the front face of the detector, which consists of a 0.6~mm layer of carbon that acts as a vacuum and thermal shield for the Si(Li) crystal inside the detector.  The crystals are kept at liquid-nitrogen (LN$_2$) temperatures for operation.  The LN$_2$ is provided to the crystals by an individual dewar directly attached to the cryostat that is controlled by the ISAC-EPICS~\cite{EPICS} system.  The detectors are structurally supported by a custom-built aluminum frame that surrounds the central plane of the EBIT, which is mounted at the base of the magnet housing.  The current mounting point of the frame has been a source of mechanical vibrations from the EBIT compressor, and is discussed in detail in Section~\ref{vibrationNOISE}.  To reduce the detection of ambient background in the Si(Li) crystals, the outer casing of each detector is surrounded radially by 2~mm of copper, followed by 1~mm of low-activity\footnote{A $^{210}$Pb activity of $\leq70$~Bq/kg.} lead, which reduces the overall ambient background contribution to the measured spectra by more than a factor of 3.

\subsection{\label{power}Power supply and conditioning}
Each detector contains a Canberra PSC 854 transistor-reset preamplifier, which provides both energy and timing outputs with a nominal impedance of 50~$\Omega$.  The low-voltage power for the preamplifier and detector electronics is provided by a DC $\pm$28~V power supply, while passively cooled linear voltage regulators provide voltages of DC $\pm$12~V and $\pm$24~V separately for each detector, with a stability of $\sim1$~mV.

The individual crystals are biased to between $-550$~V and $-600$~V using an 8-channel {\sc iseg} EHS 8210x high-precision high-voltage (HV) power supply.  The power-supply is controlled via a CAN-interface, and has an auto-shutdown feature in case of a detector warm-up.  The detector preamplifier, HV, and data acquisition (DAQ) power supplies are all protected by a UPS backup system, which provides pure sine-wave power conditioning with an output voltage regulation of $\pm2\%$.

\subsection{Electronics and signal processing}
The preamplified signals from the detector are conditioned by a custom-built signal-processing amplifier before being digitized by an analogue-to-digital converter (ADC).  The development and construction of a custom amplifier was necessary to both filter and decouple the 4~V transistor-reset output signal from the Si(Li) detector.  After processing, the signal digitization is performed by a self-triggered, 100~MHz 8-channel SIS3302 FPGA-based sampling ADC~\cite{STRUCK}.  For each channel, the trigger threshold and required pulse shape (or rise time) are set individually. The ADC hardware applies a trapezoidal energy filter to generate a moving-average window with adaptable parameters that are used to approximate the pulse integration~\cite{STRUCK}. Once triggered, each pulse is recorded with a 48-bit time-stamp generated by the ADC clock that is offset by the EBIT PPG signal.  This provides a time for each event relative to the start of each measurement cycle.

In addition to recording energies and times, the ADC records full waveforms for each signal which can be subsequently used for an off-line pulse-shape analysis (PSA). This method of analysis allows for the removal of invalid signals caused by noise and false triggers for the reduction of background events, as well as filtering and fitting valid waveforms to improve the spectral resolution.

\subsection{{\sc geant4} Simulations}
\begin{figure}[t!]
\begin{center}
\includegraphics[width=\linewidth]{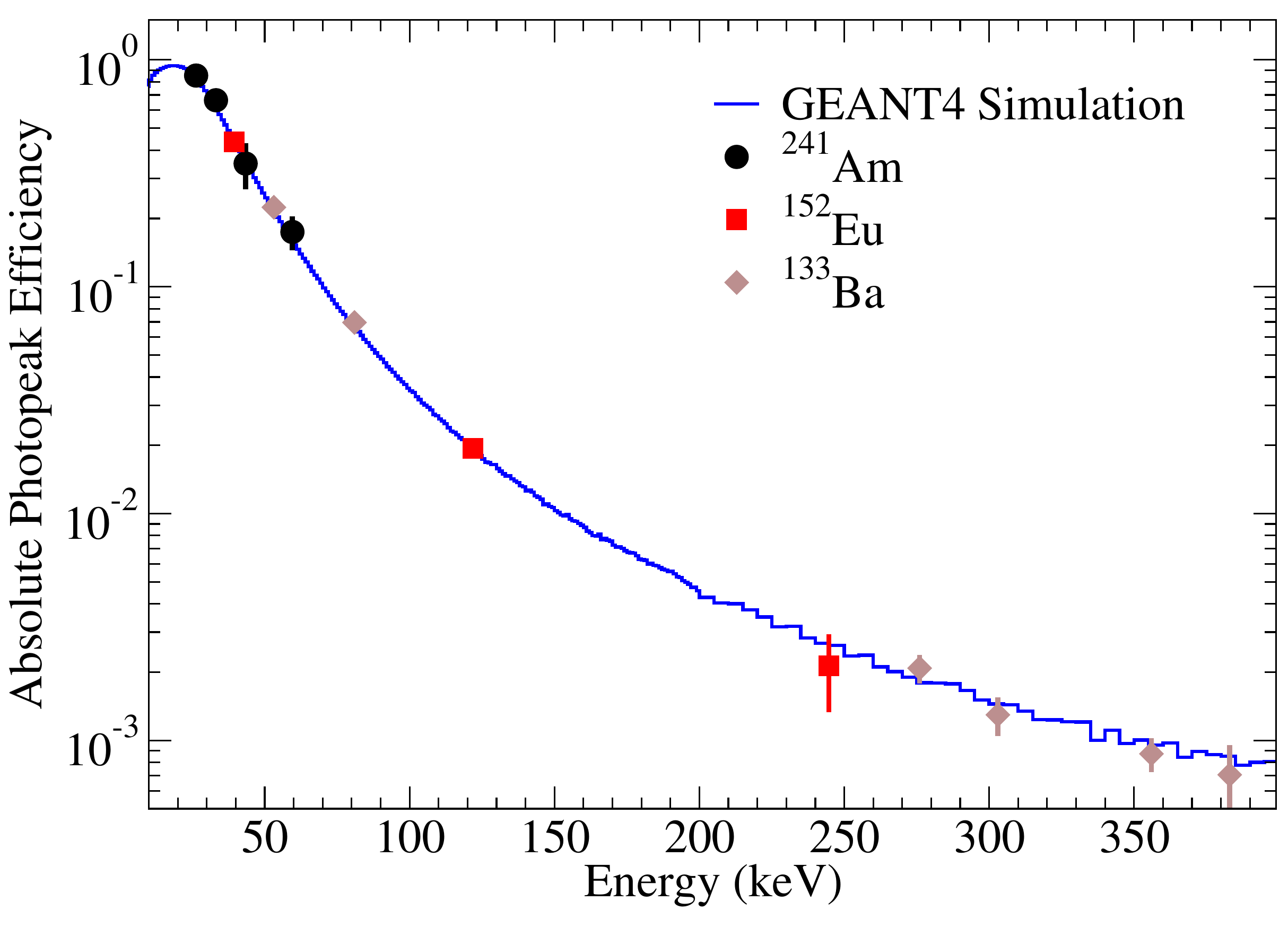}
\end{center}
\caption{\label{Efficiencies}The absolute efficiency response as a function of energy for a typical Si(Li) detector, compared to a {\sc geant4} simulation.  The experimental data were acquired using three different radioactive calibration sources when the Si(Li) detector was not mounted to the trap.  The error bars result from both the statistical and source-to-detector distance uncertainties, and in most cases are smaller than the data points.}
\end{figure}
One of the current limitations of this apparatus is a lack of access to the center of the trap for performing calibrations with radioactive sources.  Therefore, detailed simulations are required to model the detector efficiencies during on-line running conditions.  A {\sc geant4}~\cite{GEANT} simulation was therefore developed to properly model the spectral response of the Si(Li) array surrounding the EBIT.

Of particular importance for determining the detection efficiency of the apparatus, a simulation of the realistic ion-cloud distribution is required, since the solid-angle acceptance differs due to the slit sizes in the electrode-housing cylinder (Fig.~\ref{trap_interior}(a)).  This variation in the access-port geometries generates a decrease in acceptance for the small ports of nearly a factor of 2, leading to a 1.9\% geometric acceptance for the full array.

Similar to the geometric acceptance, the intrinsic response of the crystals must be accurately reproduced by the simulation in order to generate the correct absolute efficiency of the array.  This was accomplished by varying the carbon-window thickness and Si(Li) dead-layer in the {\sc geant4} geometry to match the observed crystal response from source measurements.  A comparison of the experimental and simulated photopeak detection efficiencies for one of the Si(Li) detectors is displayed in Fig.~\ref{Efficiencies}.

\section{\label{environment}Environmental Effects}
The TITAN facility is located roughly 5~metres above the floor on a raised platform in the ISAC-I experimental hall.  As a result, environmental fluctuations which may affect the sensitivity of the experiment must be continuously monitored.  To accomplish this, several diagnostic components are situated in various locations around the experimental setup, including thermocouples, vibration sensors, optical-light sensors, and voltage monitors.

\subsection{Thermal instabilities}
The day/night temperature variations in the experimental hall were observed to be between $5$ and $10^\circ$~C, with a maximum summer temperature near the EBIT approaching $35^\circ$C.  These thermal instabilities manifest themselves in gain drifts of the preamplifier electronics, which were observed to be $<1\%$ and can be corrected for.  The detector resolution and efficiencies were shown to be constant over this temperature range and are thus not affected by the thermal cycles.  The ADC and amplifier are also located in a non-temperature controlled environment, which can reach $40^\circ$C in the summer and may also contribute to the observed gain shift.  This effect can also be corrected for, and does therefore not generate anomalously poor resolutions.

\subsection{\label{vibrationNOISE}Vibration-induced noise}
\begin{figure}[t]
\begin{center}
\includegraphics[width=\linewidth]{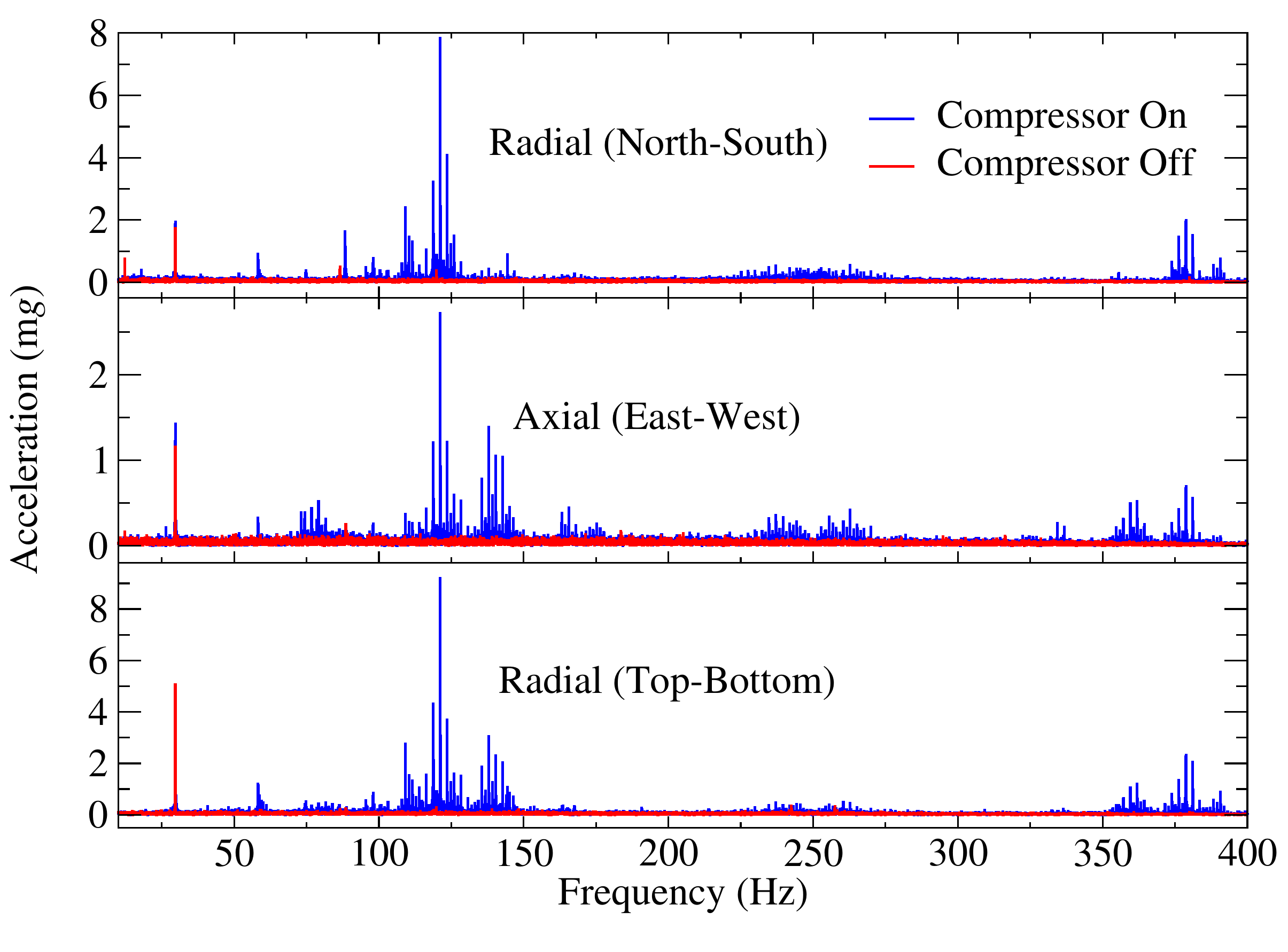}
\end{center}
\caption{\label{FFT}Observed vibrational frequencies from 10-400~Hz on the horizontal access ports at 90$^\circ$ and 270$^\circ$ when the cryo-cooler is on (blue) and off (red).  The low-frequency region has been omitted to highlight the induced vibrational resonances in the Al frame.  These measurements result from a fast-Fourier-transform (FFT) performed on data acquired by a TinkerForge inertial-measurement unit (IMU)~\cite{Tinker}.}
\end{figure}
The EBIT employs a two-stage Gifford-McMahon helium cryo-cooler that keeps the superconducting magnet at temperatures of $<6$~K using a liquid-helium-free system~\cite{Lap10}.  The cooling is performed by a compressor unit that supplies high-pressure He gas to the cold-head and re-compresses the returned gas.  In this process, the compression cylinder generates low-frequency (1.2~Hz) vibrational noise, which is subsequently transferred to the EBIT through the cold-head.  The low-frequency noise does not pose a significant concern to the extracted signal due to the filtering process that is applied before digitization.  However, these vibrations resonate at many frequencies in the aluminum detector-support frame and generate acoustic noise up to several-hundred Hz.  The distribution of high-frequency vibrations that exist at one of the horizontal access ports is displayed in Fig.~\ref{FFT}, and shows significant noise at $\sim120$~Hz and $\sim380$~Hz which is only present when the compressor is running.  The magnitude of this effect varies from detector to detector, and is correlated to the distance each port has from the frame mounting.  A decrease in resolution by more than 20\% at 50~keV results from mounting the detectors directly to the Al frame, with no isolation.  To reduce this effect, the use of vibration-isolation material is currently being implemented in the system, and is presented in detail in Ref.~\cite{Lea14b}.

\subsection{Magnetic field effects}
The radial field strength at the crystal location is roughly 5\% of the value at the trap center, as displayed in Fig.~\ref{BfieldFig}.  Thus, for a typical magnet setting of 4~T, the field experienced by the Si(Li) crystals is 0.2~T.  To confirm previous investigations of ${\vec B}$-field effects on HPGe detectors~\cite{San07,Agn09}, the effect on the Si(Li) detection efficiency, spectral resolution, and ADC channel number were investigated for fields at the trap center of 0-2.5~T, in 0.5~T steps.  These studies were performed using a $^{133}$Ba source placed on the outer housing of one of the Si(Li) detectors while it was mounted on the trap.  The spectral resolution at 53~keV was found to be constant to within $5\%$, and no variations in the detection efficiency were observed to within $1\%$.  A slight increasing linear trend in the ADC location of the peak centriod was observed as a function of photon energy for different ${\vec B}$ fields, and the dependence was found to be at most 0.27~channels/keV.  The centroid shifts do not pose a problem, as the magnetic field of the EBIT decays by less than $1.2\%$ per week, and a typical experimental run is less than two hours.

\section{\label{Commish}On-Line Commissioning}
The first on-line commissioning with radioactive beam was performed using six Si(Li) detectors, and is reported in Ref.~\cite{Len14b}.  The goal of this measurement was to characterize and examine the capabilities of the setup through the observation of X-rays resulting from $^{124}$Cs EC decay.  This case was chosen as the initial measurement due to its relatively large EC branching-ratio and short half-life, which are both well known and therefore suited to providing a benchmark test.  The summed data over $\sim48$~h for both portions of the measurement cycle from 15-130~keV are displayed in Fig.~\ref{124Cstimeslice}, highlighting the observed X- and $\gamma$-ray lines from both $^{124}$Cs and $^{124}$In.  The sections below outline the successful demonstration of the technical upgrades to the system under on-line running conditions.

\subsection{\label{storagelosses}Storage losses}
\begin{figure}[t!]
\begin{center}
\includegraphics[width=\linewidth]{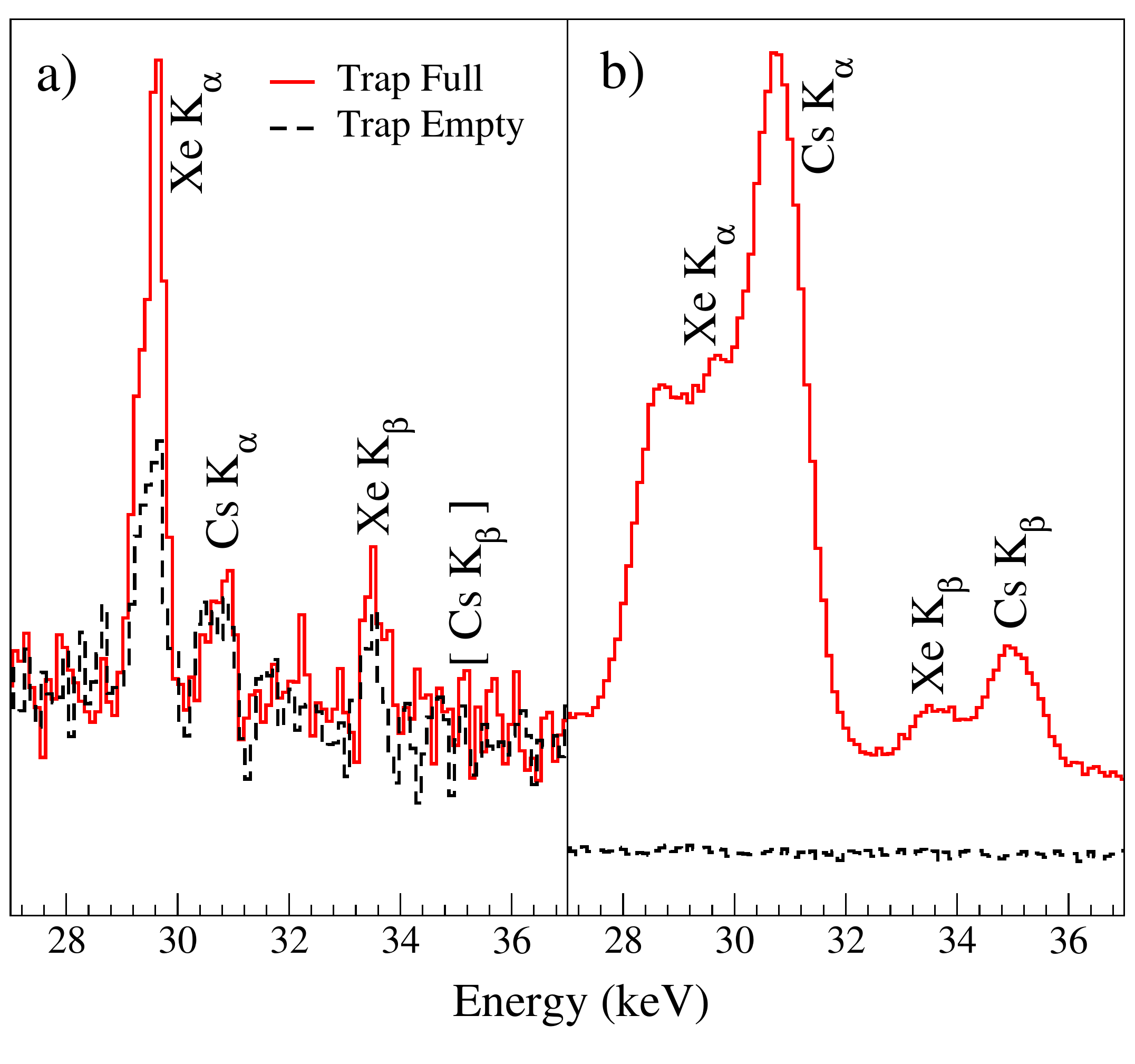}
\end{center}
\caption{\label{BrunnerCompare}A comparison of the observed photon spectrum from 27~keV to 37~keV for both the trap-full and trap-empty portions of the EBIT cycle.  The two panels compare the results when the ion-trap is operated in (a) Penning-trap mode with a LEGe detector (Ref.~\cite{Bru13}) and (b) EBIT mode with Si(Li) detectors (this work and Ref.~\cite{Len14}).  In panel (a), the X-rays result from the EC decay of $^{126}$Cs, using several hours of constant injection followed by a long trap-empty spectrum.  The presence of decay X-rays in the trap-empty spectrum demonstrates the significant loss of ions in the trap.  In panel (b), a cycled mode was used (see text) where the shorter-lived $^{124}$Cs was trapped for 20~s followed by 5~s of trap-empty data.  The differences in the intensity and resolution of the observed X-rays results from different production targets and detectors, respectively.}
\end{figure}
Previous decay measurements performed with TITAN suffered from a continual loss of ions after injection, which were likely due to off-axis injection and extraction.  A mitigation of these effects was possible in the present work by exploiting improved ion manipulation provided by the electron beam.

To investigate the improved confinement effects, significantly longer trapping times were employed using a similar RIB\footnote{The measurement in Ref.~\cite{Bru13} does not show signs of the In decay, as it used a different ISAC production target.} to Ref.~\cite{Bru13}, which primarily consisted of $^{124}$Cs ($t_{1/2}=30.8(5)$~s~\cite{NNDC}).  The RIB was delivered to the TITAN-RFQ where it was accumulated, cooled, and bunched for 1~s in the RFQ, and subsequently transported at 1.5~kV to the EBIT where it was stored for 20~s.  Following storage and decay, the ions were pulsed out of the trap, and 5~s of trap-empty background was measured to characterize possible in-trap ion losses.  The summed data over $\sim48$~h for both portions of the measurement cycle from 15-130~keV are displayed in Fig.~\ref{124Cstimeslice}, highlighting the observed X- and $\gamma$-ray lines present during the storage portion of the cycle.  The significant suppression of background radiation, as well as the non-observation of the photon lines in the trap-empty spectrum, demonstrate the complete removal of ions from the trap.  This represents a significant trapping improvement over the previous works in Refs.~\cite{Ett09,Bru13}, as demonstrated in Fig.~\ref{BrunnerCompare}.  Storage times of up to a minute were also tested, and no evidence of ion loss was observed, suggesting that even longer trapping cycle times may be employed in the future.

\subsection{Suppression of 511~keV annihilation radiation}
\begin{figure}[t]
\begin{center}
\includegraphics[width=\linewidth]{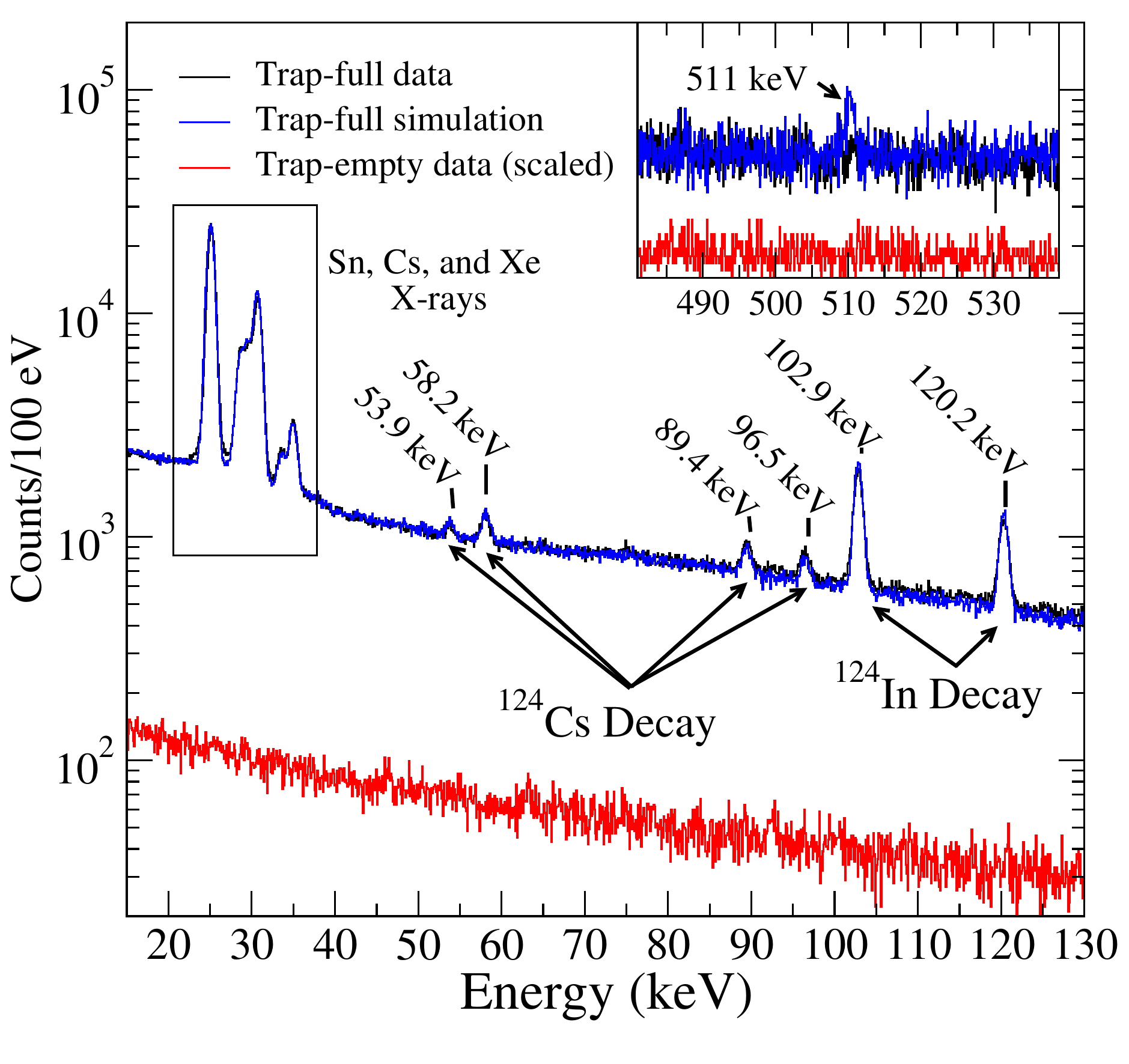}
\end{center}
\caption{\label{124Cstimeslice}The observed photon spectrum from 15-130~keV, taken during the commissioning experiment, showing both the trap-full data (20~s/cycle) and simulation in black and blue, respectively.  The trap-empty data (5~s/cycle) (red) is representative of the ambient photon background, and has been time-scaled for a direct comparison.  The complete removal of ions during the extraction phase of the cycle is demonstrated by the absence of X- and $\gamma$-ray lines in the trap-empty background spectrum.  The inset displays the energy region from 480-540~keV, highlighting the absence of the 511~keV positron annihilation radiation relative to a {\sc geant4} simulation with no $\vec{B}$ field.}
\end{figure}
The decay-particle trajectories described in Section~\ref{traj} also serve to suppress the detection of 511~keV positron annihilation radiation.  Since the decay of $^{124}$Cs has a significant $\beta^+$-decay branch, a direct comparison of the efficiency-corrected photopeak areas at 354~keV and 511~keV provide an estimate for the positron-annihilation background suppression.  The suppression factor
\begin{eqnarray}
\label{suppressionfact}
S_{511}&=&\left(\frac{I_{511}}{I_{354}}\right)_{\rm lit.}\times\left(\frac{N_{354}}{N_{511}}\right)_{\rm exp.}\times\left(\frac{\epsilon_{511}}{\epsilon_{354}}\right)_{\rm sim.},
\end{eqnarray}
\noindent
where $I$ is the relative peak intensity from the literature~\cite{NNDC}, and $\epsilon$ is the simulated absolute photo-peak detection efficiency. The ratio $\frac{N_{354}}{N_{511}}$ is the fraction of observed counts in the 354~keV peak relative to 511~keV.  However, the data does not display any evidence of a 511~keV peak (Fig.~\ref{124Cstimeslice} (inset)), the relative peak area used in Eqn.~\ref{suppressionfact} represents the statistical $1\sigma$ upper limit of 26~counts above ambient background.  A lower limit on the suppression effect was thus found to be a factor of 20, which was also validated through a comparison of the experimental data to a {\sc geant4} simulation with no magnetic field (Fig.~\ref{124Cstimeslice} (inset)).  This comparison also serves as an analogue estimate for $\beta^-$ removal from the trap.

\subsection{\label{HCIEff}Atomic-structure effects}
\begin{figure}[t!]
\begin{center}
\includegraphics[width=\linewidth]{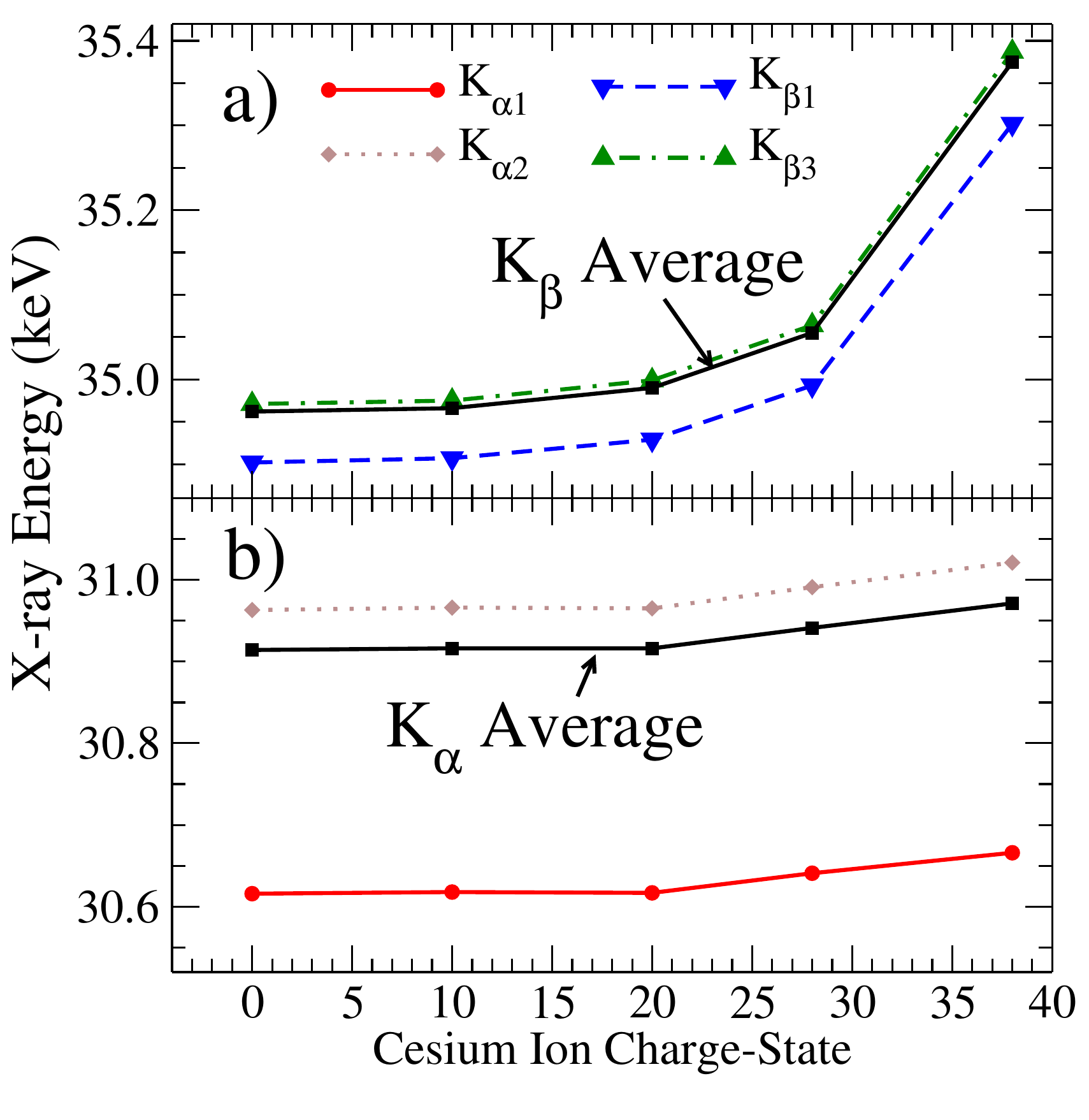}
\end{center}
\caption{\label{Shifts}The theoretical change in a) $K_\beta$ and b) $K_\alpha$ X-ray energies as a function of charge-state for Cs ions.  The calculations were performed using the multiconfigurational, relativistic Dirac-Fock software, {\sc fac}~\cite{FAC}, and display increasing $K_\alpha$ and $K_\beta$ energies as the charge-state increases.  The dashed lines serve to illustrate the increasing energy difference relative to a neutral Cs atom.  The weighted-average energy curves for $K_\alpha$ and $K_\beta$ result from the relative X-ray intensities, and represent what would be observed experimentally due to limited energy resolution.}
\end{figure}
In addition to the significant trapping advantages provided by the electron beam, the atomic structure of the trapped ions are also altered through electron-impact ionization~\cite{HCI} (Fig.~\ref{Shifts}).  This effect was observed in the commissioning experiment, and the observed relative average X-ray energy shift for $^{124}$Xe of $\langle K_\beta-K_\alpha\rangle=90(40)$~eV was in good agreement with the calculated value of 92~eV~\cite{Len14b}.  Furthermore, no $K_{\beta2}$ X-rays were observed since the charge states of the trapped Cs ions ($\approx28^+$) corresponded to a fully stripped $N$-shell.  A distribution of ions in various charge-states simultaneously exist in the trap, which is well understood and has been investigated previously for TITAN's EBIT~\cite{Lap10}.  Additional atomic effects were also observed, including changes in the $K_\alpha/K_\beta$ ratio, and in all cases were found to be consistent with theoretical calculations~\cite{Len14b}.

\subsection{\label{trapSIMcompare}Comparison to simulation}
Benchmarking of the {\sc geant4} simulations to in-trap data was also performed for the commissioning experiment.  A comparison of the simulated and measured spectra is displayed in Fig.~\ref{124Cstimeslice}.  Since the exact in-trap contributions of each species can vary from experiment to experiment, these quantities must be derived from the observed spectra.  As a result, the simulated decay spectra are individually scaled and combined with an ambient-background spline function derived from the trap-empty measurements.  The total Monte-Carlo spectrum (blue) that results from this procedure exhibits general agreement with the data.  The photon-energy deposition in the crystal is combined with the realistic Si(Li) response function that is derived from calibration measurements.  These response functions include crystal imperfections that lead to slightly asymmetric photo-peaks due to incomplete charge-collection and trapped-charge effects~\cite{KNOLL}.

\section{Future Upgrades}
\subsection{\label{MI}Multiple ion-bunch stacking}
For RIBs from ISAC with yields $\geq10^6$~s$^{-1}$, the limiting factor for ion-storage in the EBIT is the space-charge of the RFQ ($\sim10^5$-$10^6$)~\cite{Bru12}.  For nuclei with small branching ratios ($<10^{-4}$), this limit would exclude the possibility of performing statistically significant measurements within a reasonable amount of time.  As a result, a method for overcoming this space-charge limit was recently tested using a beam of $^{116g,m}$In, with short ($\approx25$~ms) RFQ accumulation times, and the subsequent injection of many ion bunches into the EBIT without extraction~\cite{Lea14}.  By using this multiple-injection technique~\cite{Ros13}, it was possible to stack several hundred ion-bunches in the EBIT, thus allowing for significantly more ions to be stored in the trap for one decay cycle.  The demonstration of this technique has opened the venue for experiments that were previously unfeasible.

\subsection{Isobaric purification with the MR-ToF technique}
One of the advantages of manipulating ion-bunches in a multi-trap system is the possibility of isobaric cleaning.  This form of beam purification has traditionally been performed with the assistance of a buffer-gas filled Penning trap, however this technique typically limits the total number of charges allowed to $\sim10^3-10^4$~\cite{Sav91}.  The TITAN facility is currently in the process of implementing a multi-reflection time-of-flight (MR-ToF) device\footnote{This device was designed and constructed in Germany at JLU Gie\ss{}en~\cite{Pla13} and is currently undergoing offline commissioning at TRIUMF.}, which can achieve ion capacities in excess of $10^{6}$ ions per second, while maintaining a mass resolving power of $\Delta m/m\geq10^{5}$~\cite{Pla13}.  This component is to be included downstream from the RFQ, thus allowing for purification of SCIs before they are injected into the EBIT.

\subsection{High-purity germanium detectors}
\begin{figure}[t]
\begin{center}
\includegraphics[width=\linewidth]{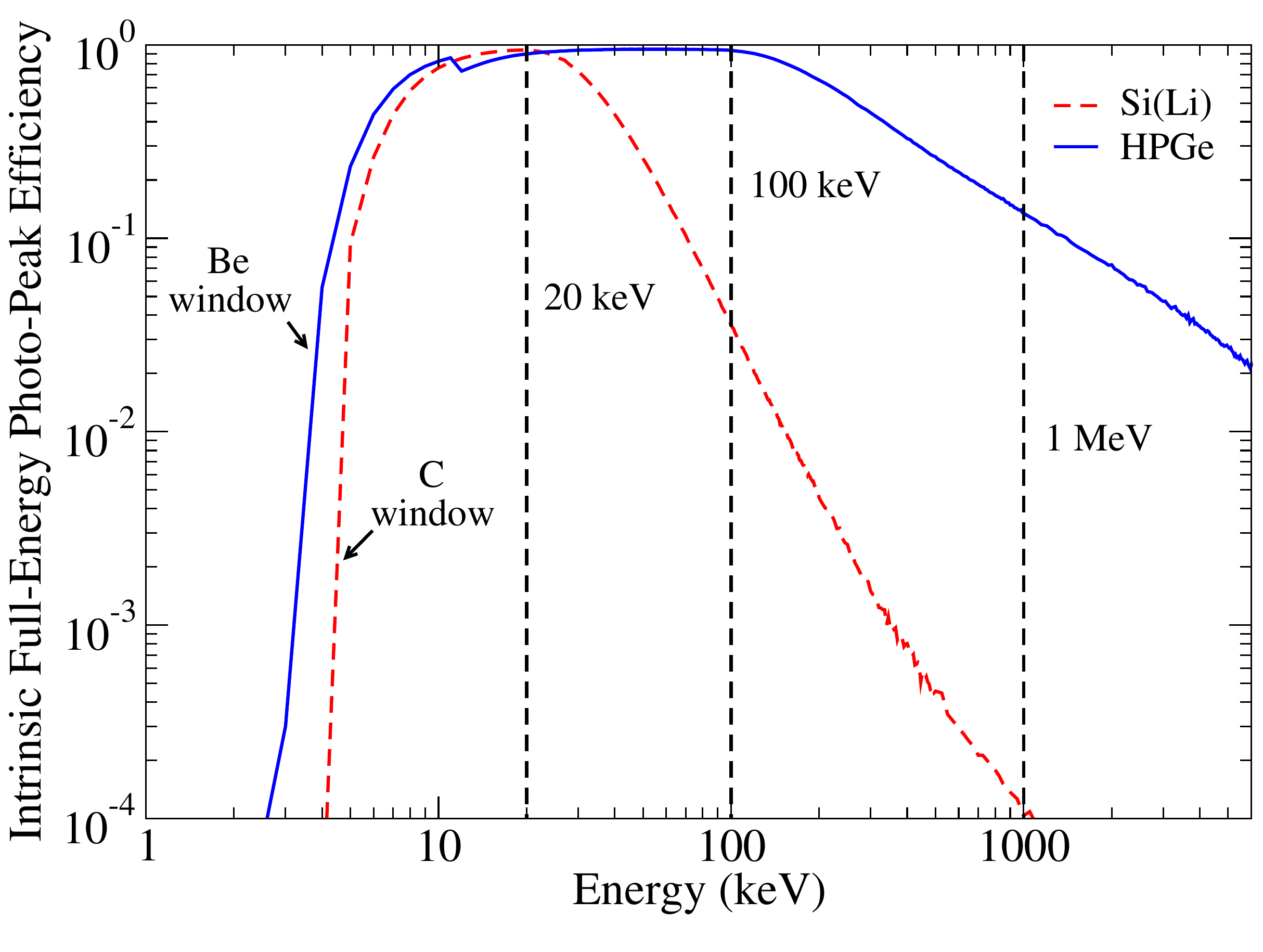}
\end{center}
\caption{\label{SiLivsHPGe}A comparison of the simulated intrinsic efficiencies for a TITAN Si(Li) detector and an 8$\pi$ HPGe detector separated from the source by the EBIT Be windows.  The efficiency profiles are nearly identical below $\sim20$~keV, but a large increase in detection efficiency above this can be gained from using HPGe.  This energy-dependent response therefore increases the versatility of performing decay spectroscopy in-trap with the TITAN EBIT.  The low-energy character of the respective curves is determined by the thickness of the Be and C front-face windows on the HPGe and Si(Li) detectors, respectively.}
\end{figure}
Although the current photon detectors have a high low-energy photon detection efficiency, it drops dramatically at roughly $30$~keV which limits the range of experiments that are possible.  The experimental capabilities can be increased using HPGe detectors, thereby increasing the versatility of performing in-trap decay spectroscopy with TITAN.  With the recent decommissioning of the 8$\pi$ $\gamma$-ray spectrometer~\cite{Sve03} from TRIUMF, the prospect of deploying up to seven of these detectors in the ports around the EBIT exists due to their compatible size and availability.  Each individual detector is composed of a cylindrical HPGe crystal with a radius of 2.65~cm, and a length of 6.0~cm.  The crystals are located in an LN$_2$-cooled cryostat, which is heat- and vacuum-shielded by a thin Be window on the front-face.  This set-up has already been modelled in a {\sc geant4} simulation, and a comparison of the simulated absolute efficiencies for the HPGe and Si(Li) crystals are displayed in Fig.~\ref{SiLivsHPGe}.

\section{Summary and Conclusions}
In summary, a significantly improved in-trap decay spectroscopy setup has been developed using the TITAN facility at TRIUMF.  The apparatus consists of 7 low-energy planar Si(Li) detectors which surround the TITAN EBIT; an open-access charge-breeding ion trap with a magnetic field of up to 6~T.  The current goal of this new facility is to provide a low-background environment for the observation of weak EC branching ratios of the intermediate nuclei for $\beta\beta$ decay.  The ion-trap environment allows for the detection of low-energy photons by providing backing-free storage, while simultaneously guiding charged decay particles away from the trap center via the strong magnetic field.  When combined with the intense electron beam of the EBIT, the strong magnetic field provides excellent ion confinement, which allows for storage times of minutes, or more.  Impact ionization induced by the electron beam increases the typical charge-states of the trapped ions to such a level that changes to the atomic structures were observed via X-ray energy shifts and $K_\alpha/K_\beta$ ratio changes.  Although these atomic-structure alterations are a byproduct of the improved ion storage, these effects could be exploited in the future for studies on $\beta$ decay of HCIs~\cite{Lit11}.  The background reduction provided by the apparatus presented in this work represents a significant step towards measuring weak branching ratios of $10^{-4}$ or less.  This new facility is therefore poised to make a significant impact in the field of low-intensity spectroscopy.

\section{Acknowledgements}
TRIUMF receives federal funding via a contribution agreement with the National Research Council of Canada (NRC).  This work was partially supported by the Natural Sciences and Engineering Research Council of Canada (NSERC), and the Deutsche Forschungsgemeinschaft (DFG) under grant FR 601/3-1.  TDM and ATG acknowledge support from the NSERC PGS-M and CGS-D programs, respectively.  KGL would like to thank Stephan Ettenauer for many useful discussions on this work and its history with TITAN.  The authors also thank Dave Morris, Chris Pearson, Leonid Kurchaninov, Fuluny Jang for their many contributions to this system.

\end{document}